\documentclass[twocolumn,pre,10pt,amsmath,amssymb,nofootinbib,showpacs,superscriptaddress,floatfix]{revtex4}

\newcommand{\dfr}[2]{\frac {\displaystyle #1}{\displaystyle #2}}
\usepackage{graphicx}
\usepackage[hypertex]{hyperref}
\begin{document}
\author{M.~G.~Vasin}
\title{Gauge theory description of glass transition}
\affiliation{Physical-Technical Institute, Ural Branch of Russian Academy of Sciences, 426000 Izhevsk, Russia}
\affiliation{Institute of High Pressure Physics, Russian Academy of Sciences, Troitsk 142190, Moscow Region, Russia}
\begin{abstract}
An analytical approach, which develops the gauge model of the glass transition phenomenon, is suggested. It is based on the quantum field theory and critical dynamics methods. The suggested mechanism of glass transition is based on the interaction of the local magnetization field with the massive gauge field, which describes frustration-induced plastic deformation. The example of the three-dimensional Heisenberg model with trapped disorder is considered.
It is shown that the glass transition appears when the fluctuations scale reaches the frustrations scale, and the mass of the gauge field becomes equal to zero. The Vogel-Fulcher-Tammann relation for the glass transition kinetics and critical exponent for non-linear susceptibility, $1.7\lesssim \gamma < 3$, are derived in the framework of the suggested approach.

\end{abstract}

\maketitle

The formulation of a universal theory of glass transition has remained one of the most intriguing but still unresolved problems of condensed matter physics~\cite{N1, N2, Tanaka} for a long time. Many systems which manifest this phenomenon regardless of their nature enable us to conclude that this phenomenon does not depend on any microscopic details, but is determined by the symmetry properties of the systems, as in the case of phase transitions, regardless of the scale.
However, it is well known that the glass transition is not accompanied by divergence of the static correlation length of the local magnetization and linear susceptibility, $\chi_L$, typical for critical phenomena. This does not allow to assign this glass transition phenomenon to phase transition.
To support the opinion that glass transition is a critical phenomenon, it should be noted that critical slowing down of all processes takes place at the microscopic level during glass transition and the nonlinear susceptibility, $\chi_N$, diverges. Besides, recent computer simulations enable us to suggest that glass transition is still a specific form of critical phenomena, in which the scale of a temporary ordered region of the medium-range crystalline order is considered to be a diverging static correlation length~\cite{Tanaka}.
These are the features of a critical phenomenon, however, the reason for this still remains unclear.

The theoretical approach to the glass transition description, introduced in this paper, is based on well known methods of the critical phenomena description and nonequilibrium dynamics, and promotes the ideas mentioned in \cite{Toul, Hertz, Volovik, Kanazawa, Riv, Nelson, Kivelson, Nusinov}. According to this theory the glass transition is accompanied by the formation of local ordering in the same way as the second order phase transition. We suppose that the key feature, which distinguishes one phenomenon from the other, is the presence of the disorder-induced frustrations. The frustrations can be induced by both quenched randomness, associated with the random sign of the spin interaction in the case of the spin-glass transition, and topological properties of the system, as the geometrical frustration in~\cite{Kivelson}. The frustrations distort the system's space \cite{Riv}, make it heterogeneous, and lead to its plastic deformation. We describe these frustrations below as topological defects (disclinations), which are sources of appropriate gauge fields \cite{Hertz, Volovik, Riv, Nelson, Kivelson, Nusinov, Kadich}. The fluctuations of the local magnetization field in the disordered phase increase as the temperature approaches the phase transition point, $T_c$. However, the system contains the frustrations, whereas the ordered domains should not contain them. Therefore, the fluctuations grow until their size remains less than the typical spacing between the frustrations, $L_{fluct}\ll L_{frust}$. Since the fluctuations can not grow more than $L_{frust}$, the critical slowing-down occurs while they reach this size. Below we show that this critical behavior is attributed to vanishing the effective mass of the gauge field, which happens at the temperature $T_g>T_c$. As a result, the system freezes in a state with a disordered structure. Thus, the glass transition represents the critical slowing-down of fluctuations growth by disorder-induced frustrations, that complies with the ``frustration-limited domain theory'' (FLDT) \cite{Kivelson} and correlates with other approaches \cite{Riv, Rizhev}.

In more detail, let us consider an example of the three-dimensional static isotropic Heisenberg model, whose order parameter, ${\bf s}$, is considered to be a local magnetization vector with the SO(3) continuous symmetry group. The pure model describes the system undergoing the paramagnetic--ferromagnetic phase transition, with no quenched disorder available. The static action of the model is well known and has the form of
\begin{gather}
    S=\int \left[\dfr 12(\partial_i {\bf s })^2+U({\bf s})\right]d{\bf r},
\end{gather}
where $U({\bf s})=\mu^2 {\bf s}^2 +v{\bf s}^4 $, .
Note that at $\mu ^2<0$ the action is invariant under the SO$(3)$ gauge transformations and $\langle{\bf s}\rangle =0$. However, at $\mu ^2<0$ the symmetry is explicitly broken, since the system can arbitrarily ``choose'' only one state from all equivalent states with the minimum energy $U({\bf s})$ potential minima situated on the $|{\bf s}|=i\mu/\sqrt{2v}$ sphere. We fix the vacuum by means of fixing a point on the sphere. The system is no longer symmetrical with respect to the SO$(3)$ gauge group, but it is invariant under the SO$(2)$ group of the rotation around the chosen direction.

In a disordered spin system the equilibrium spin orientations at different points are not collinear and a suitable connection between orientations  is done by introducing a gauge field, $A^{a}_{\mu }$, and replacing the ordinary derivative, $\partial_i{\bf s}$, by a covariant derivative, $D_i{\bf s}$, \cite{Riv, Kadich}:
\begin{gather}\label{1}
    S=\int\left[\dfr 12(D_i {\bf s })^2+U({\bf s})+\dfr 14F_{\mu\nu}^aF^{a}_{\mu\nu}+J_{\mu }^aA^{a}_{\mu }\right]d{\bf r},
\end{gather}
where
\begin{gather}\label{3}
D_i s^a=\partial_{i}s^a+g\varepsilon^{abc} A_{i}^bs^c, \\ F_{\mu\nu}^a=\partial_{\mu}A_{\nu}^a-\partial_{\nu}A_{\mu}^a+g\varepsilon^{abc} A_{\mu}^bA_{\nu}^c,
\end{gather}
$J_{\mu }^a$ is the source of the $A_{\mu}^a$ field which is introduced to the action in addition to the general part. The reason for this is discussed below.

The method of introduction of disorder in the theory is very important and plays a key part.
In order that a spin system possess glass properties we should inject to this system a disorder which gives rise to frustration of its structure. Unfortunately, before ones did not pay attention to this fact properly.
In \cite{Volovik, Kanazawa}, for example, the gauge field is free, and any quenched randomness is absent.
This is right when the dynamic soliton solutions appear spontaneously in the form of pair of oppositely directed vortexes. However, in our case the frustrations should be quenched and static.
In \cite{Hertz} and \cite{Riv} the disorder is associated directly with the quenched gauge field, and the $A_{\mu }^a$ field is frozen in an arbitrary configuration with some $P(A_{\mu }^a)$ distribution function. It is supposed that the quenched gauge field describes the frustrations.
But it is not quite right too, since presence of the quenched gauge field in the system does not mean yet presence of frustrations.
The quenched frustrations can be represented by immovable disclinations passing through the frustration planes~\cite{Riv, Nusinov} (see Fig.\,1). They are the static sources of the $A_{\mu }^a$ gauge field, and should be injected in the model by means of the static sources field, $J_{\mu }^a$.
Therefore, in contrast to \cite{Hertz, Riv}, we believe that it is more correct to consider the source field, $J^{a}_{\mu }$, but not the gauge field, $A_{\mu }^a$, in the capacity of quenched random field. In this case the $A_{\mu }^a$ field remains to be dynamic one. For illustration the $A_{\mu }^a$ field can be interpreted as a local relative rotation of neighboring spins, which corresponds to their local equilibrium. Spins can be movable, but their local equilibrium configuration around a static frustration in the $\delta V$ volume, bounded by the $\delta S$ sphere, should satisfy to the following condition:
\begin{equation}\label{Sourse}
    \dfr 12\oint\limits_{\delta S} F_{\mu\nu}^adS_{\nu }=\int\limits_{\delta V} J_{\mu}^ad{\bf r}\neq 0,
\end{equation}
which follows from the principle of least action.

Expansion of the local magnetization field, ${\bf s}$, near one of the vacuum states, for instance $\left<{\bf s}\right>_0=(0,\,0,\,i\mu/\sqrt{2v})$, in small $\phi =s-i\mu/\sqrt{2v}$ deviations, and use of the gauge transformation properties allow to rewrite (\ref{1}) in the form of the action of two massive vector bosons,  $A_{\mu }^{\kappa }$ ($\kappa =\{1,\,2\}$), with the mass $M_0=ig\mu/\sqrt{2v}$, one massless vector boson, $A_{\mu }^3$, and one scalar field, $\phi $:
\begin{multline}\label{41}
 S=\int\left[\dfr 12(\partial_{\mu } \phi )^2+2\mu^2\phi^2+\dfr{g^2\mu^2}{4v}A_{\mu }^{\kappa}A^{\kappa}_{\mu }+\dfr 14F_{\mu\nu}^aF^{a}_{\mu\nu}\right.\\
 \left.+v\phi^4+\dfr{g^2}2\phi^2A_{\mu }^aA^{a}_{\mu }+J^{a}_{\mu } A_{\mu }^a\right]d{\bf r}.
\end{multline}
Note, that this Lagrangian is not gauge-invariant in presence of external sources, that is important below.

Let us consider the system with randomly quenched frustrations, which are described by means of the sources field, $J^a_{\mu }$.
The $J^a_{\mu }$ field is static and absolutely random. Therefore, for simplicity, we assume that $\langle J^a_{\mu }({\bf r})J_{\mu }^a({\bf r'})\rangle =I_0\delta ({\bf r}-{\bf r'})$, where $I_0$ is the intensity of the quenched disorder. It is shown below that this parameter is proportional to some ``structural temperature''.
Averaging over $J^a_{\mu }$ leads to redefinition of the partition function:
\begin{gather}\label{SS}
    Z=\int \left[\int  \exp\left(-S-\dfr 14\int I_0^{-1}{J^a_{\mu }}^2 d{\bf r}\right) DJ^a_{\mu }\right]D\phi DA_{\mu }^a,
\end{gather}
where $\int \dots Dx$ is the continual integral.
It leads to additional contribution to the $A^a_{\mu}$ ``mass'', which takes the form of
\begin{gather}\label{random2}
    M^2=-I_0 +\mu^2g^2/4v.
\end{gather}
Thus, the frustrations lead to the renormalization of the gauge field mass.

The renormalization of the gauge field mass affects the critical behavior of the system, since it shifts $M^2=0$ singularity to the temperature range above the paramagnetic-ferromagnetic transition point, $T_c$.
If we assume that $\mu^2=\alpha k_B(T-T_c)$, where $\alpha $ is some constant, then from (\ref{random2}) we have the critical divergence of the $A_{\mu }^a$ field correlation radius at $T_g=T_c+4I_0v/\alpha k_Bg^2$ in the paramagnetic phase. One can suppose that this can lead to critical slowing-downs of the fluctuations.
Thus, the disorder-induced frustrations inhibit the growth of the $\phi$ field correlation length, and the system freezes in a disordered state, that conforms with the glass transition description in FLDT.

In order to examine the above assumption we should investigate the kinetics and susceptibility of the considered model near $T_g$. To do this, we will consider the non-equilibrium dynamics of the system near the critical point.
This is a classical problem. Therefore, one can apply either the method of the dynamic generating functional~\cite{CritDyn}, or the classical limit of the Keldysh technique~\cite{Kamenev}, since both these methods coincide in the classical limit.
We will choose units such that $k_{B}T_{g}=1$.  The usage of the functional technique for the description of non-equilibrium dynamics leads to the representation of the partition function of the system in the form of
\begin{equation}\label{L2}
\displaystyle Z=\int \exp (-S^*)D\vec\phi D\vec A_{\mu}^a,
\end{equation}
where
\begin{multline}\label{L3}
S^*=\frac 12\int \left[\vec\phi(t,\,{\bf r})\hat G^{-1}(t-t',\,{\bf r-r'})\vec\phi(t',\,{\bf r'})\right. \\
   \displaystyle \left.  +\vec A_{\mu }^a(t,\,{\bf r})\hat\Delta_{\mu\nu}^{-1}(t-t',\,{\bf r-r'})\vec A_{\nu}^a(t',\,{\bf r'})\right]d{\bf r}d{\bf r'}dt dt' \\
    +\displaystyle \int\left[ g\varepsilon^{abc}(\partial_{\mu}\bar A^a_{\nu })A^b_{\mu}A^c_{\nu}  +g\varepsilon^{abc}(\partial_{\mu}A^a_{\nu })\bar A^b_{\mu}A^c_{\nu}\right.\\[4pt]
   \displaystyle +g\varepsilon^{abc}(\partial_{\mu}A^a_{\nu })A^b_{\mu}\bar A^c_{\nu}
   +g^2\varepsilon^{abc}\varepsilon^{aij}\bar A^b_{\mu }A^c_{\nu}A^i_{\mu}A^j_{\nu}
   \\[8pt]
   \displaystyle \left. +g^2 \bar A_{\mu}^aA_{\mu}^a\phi^2+g^2 (A_{\mu}^a)^2\bar\phi\phi +
    v4\, \bar\phi\phi^3\right]d{\bf r}dt,
\end{multline}
$\vec\phi=\left\{ \bar\phi ,\,\phi  \right\}$, and $\vec A^a_{\mu}=\left\{ \bar A^a_{\mu},\,A^a_{\mu} \right\}$ are vectors, the components of which are named ``quantum'' and ``classical'' respectively~\cite{Kamenev},
$G^{-1}$ and $\Delta_{\mu\nu}^{-1}$ are matrices, inverse to the Green functions matrices having the following form:
\begin{equation}\label{eq:G0-1}
    \hat G=\begin{pmatrix}
                            G^K & G^A \\
                            G^R & 0 \\
                          \end{pmatrix}, \quad
    \hat \Delta_{\mu\nu}=\begin{pmatrix}
                            \Delta ^K_{\mu\nu} & \Delta ^A_{\mu\nu} \\
                            \Delta ^R_{\mu\nu} & 0 \\
                          \end{pmatrix}.
\end{equation}
Note that opposed to Yang-Mills theory here there is no necessity to introduce an additional ghost field in the theory because of the gauge symmetry breaking and the presence of the gauge field mass.
All Green functions of the theory are explicitly determined by the same reason. The components of the Green function of the scalar field are:
\begin{gather}\label{7}
   G^{R(A)} (k,\,\omega )
   =\dfr{1}{k^2+\mu^2\pm i\Gamma_{\phi}\omega },\\
   G^{K} (k,\,\omega )=\dfr{2\Gamma_{\phi}}{(k^2+\mu^2)^2+\Gamma_{\phi}^2\omega^2 },
\end{gather}
where $\Gamma_{\phi} $ is the kinetic coefficient, which corresponds to the local magnetization.
The components of the Green function of the massive gauge field are:
\begin{gather}\label{8}
    \Delta^{R(A)} _{\mu\nu}(k,\,\omega )=\dfr{\delta_{\mu\nu}}{k^2+M^2\pm i\Gamma_{A}\omega },\\[12pt]
    \Delta^{K} _{\mu\nu}(k,\,\omega )=\dfr{2\Gamma_{A} \delta_{\mu\nu}}{(k^2+M^2)^2+\Gamma_{A}^2\omega^2 },
\end{gather}
where $\Gamma_A $ is the kinetic coefficient, which corresponds to the gauge field.
Their graphic forms are presented in Fig.\,2
\begin{figure}[h]
\label{fig0}
   \centering
   \includegraphics[scale=0.45]{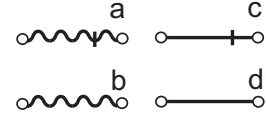}
   \caption{a) is the retard part of $A_{\mu \nu}$ field Green function , $\Delta^{R}_{\mu \nu}$; b) is the Keldysh part of this Green function, $\Delta^{K}_{\mu \nu}$; c) is the retard part of the Green function of the scalar field $\phi $, $G^{R}$; d) is the Keldysh part of this function, $G^{K}$.}
\end{figure}

The interaction of the local magnetization fluctuations per gauge field plays the key part in the considered theory (Fig.\,2).
\begin{figure}[h]
\label{fig1}
   \centering
   \includegraphics[scale=0.45]{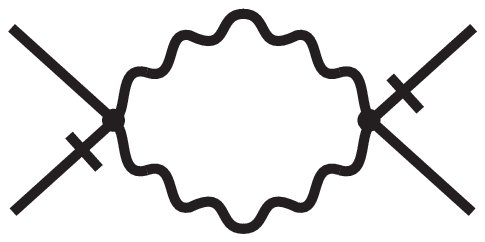}
   \caption{}
\end{figure}
This becomes clear in the process of renormalization.
Close to $M^2=0$ the gauge field becomes massless, but the scalar field remains massive with $\mu^2=4I_0v/g^2$. Therefore, the contribution to the renormalization is made only by the loops of the gauge field propagators.
According to the separation of massive field theorem \cite{Collins} the Feynman diagrams, containing the propagators of field, the mass of which is appreciably larger than the external momentum, are inversely proportional to the degree of this mass, and make a finite contribution to the renormalization.
In Fig.\,3 some graphs giving logarithmically divergent contributions to the renormalized theory are  presented. It is easy to check that near $T_g$ the considered theory is renormalizable, since the logarithmically diverging corrections lead to the renormalization of only existing terms of the Lagrangian.
\begin{figure}[h]
\label{fig2}
   \centering
   \includegraphics[scale=0.45]{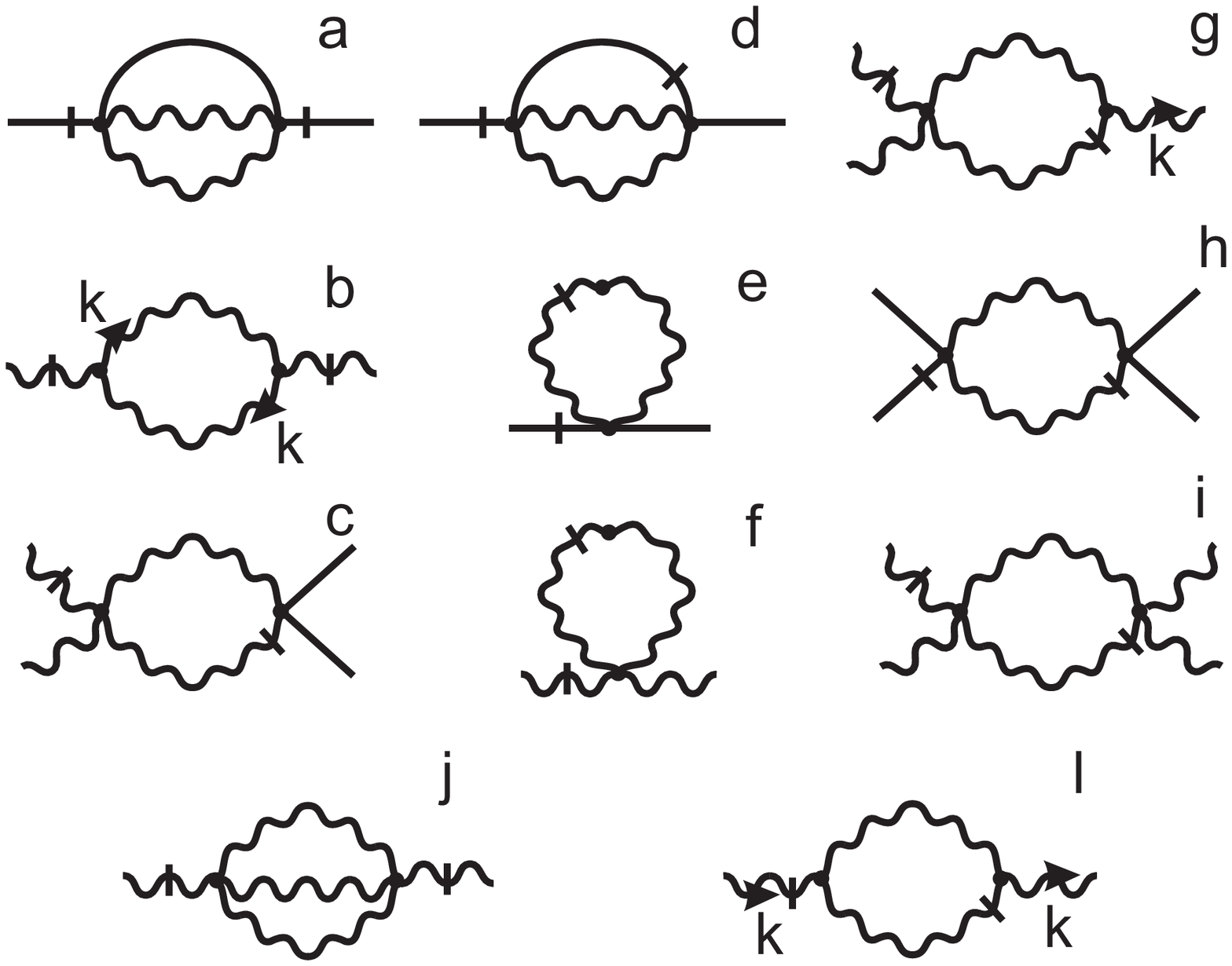}
   \caption{}
\end{figure}
However, it is necessary to consider in detail the contribution to the renormalization of the $\Gamma_{\phi}$ node of the $a$ diagram, which is shown in Fig.\,3. This term is interesting because the $\Gamma_{\phi}$ node is proportional to the relaxation time of the local magnetization field, and determines the kinetics of the glass-transition: the loop of the light field, $A^a_{\mu }$, which is given in Fig.\,2, makes the logarithmically divergent contribution $\sim \ln(1/\Lambda )\delta(\omega )$, where $\Lambda $ is the regularization parameter (cutoff of momentum). This term determines the divergence contribution of the $a$ graph (Fig.\,3) in which the massive field loop leads only to multiplying the logarithm by $4I_0v/g^2$ factor:
\begin{multline}\label{9}
    \Sigma \approx \dfr{4g^2I_0v\ln (1/\Lambda)}{\pi^2}\int\limits_0^{t_o}e^{-4I_0v|t|/\Gamma_{\phi} g^2}dt=\\
    =\Gamma_{\phi} \dfr{g^4 \ln (1/\Lambda)}{\pi^2}(1-e^{-4I_0v|t_0|/\Gamma_{\phi} g^2}),
\end{multline}
where $t_o$ is the time of the observation of the system.
One can see that for $I_0\to 0$, or with a short observation time, $\Gamma_{\phi} g^2/4I_0v\gg t_o$, this contribution becomes negligibly small, and the theory becomes nonrenormalizable. This relates to the symmetry properties of the Yang-Mills model. In this case the fluctuation dissipation theorem (FDT) is always broken because of free energy transfer between the local magnetization modes and massless gauge field modes (Goldstone modes).  However, there is also some problem in the presence of the quenched disorder: this node becomes nonlocal in time, which violates the correctness of the renormalization procedure. In order to avoid this problem it is possible to divide the time space of the system into two intervals: when $t_o\ll \Gamma_{\phi} g^2/4I_0v$ this contribution is negligibly small; but when $t_o\gg \Gamma_{\phi} g^2/4I_0v$ this graph makes the  logarithmically divergent contribution to the $\Gamma_{\phi }$ renormalization.
Hence in the one-loop approximation the renormalization group has the form:
\begin{gather}\label{RG}
   \left\{
   \begin{array}{lcl}
   \dfr{\partial \ln(\Gamma_{\phi} )}{\partial \xi}={g^4 }/{\pi^2} & \mbox{for}& t_o\gg \Gamma_{\phi} g^2/4I_0v ,\\
   \dfr{\partial \ln(\Gamma_{\phi} )}{\partial \xi}= 0 & \mbox{for}& t_o\ll \Gamma_{\phi} g^2/4I_0v ,
   \end{array}
   \right.
   \\
   \dfr{\partial \ln(\Gamma_A )}{\partial \xi}=-\varepsilon +3g^4/\pi^2 +g^2/2\pi^2 ,\\
   \dfr{\partial \ln(M^2)}{\partial \xi}= 2-3g^2/2\pi^2,\\
   \dfr{\partial \ln(\mu^2)}{\partial \xi}=2-\dfr{M^2g^2}{2\mu^2\pi^2}\approx 2 ,\\
   \dfr{\partial \ln(g^2)}{\partial \xi}= \varepsilon -g^2/\pi^2,\\
   \dfr{\partial \ln v}{\partial \xi}= \varepsilon -g^4/2v\pi^2.
\end{gather}
where $\xi=\ln(1/\Lambda )$, $\varepsilon = 4-d$. Actually, this form of the renormalization group takes into account the FDT violation on small time scales, and seems to be natural for the quasiergodic system.
In \cite{Vasin} it was shown, that matching the marginal solutions in (\ref{RG}) leads to the Vogel-Fulcher-Tammann relation for the temperature dependence of the system relaxation time near $T_g$. From the condition of the stable point existence, ${\partial \ln (g^2)}/{\partial \xi}=0$, ${\partial \ln (v)}/{\partial \xi}=0$, we get
$g^2=\pi^2\varepsilon$, $v=g^2/2$, and, using the result of~\cite{Vasin},
\begin{gather}\label{VFT}
    \tau_{rel}=\Gamma_{\phi} \propto\exp\left(\dfr{2vg^4T_g}{\alpha\pi^2(T-T_g)}\right).
\end{gather}
Hence, the critical slowing down of all relaxation processes does occur in $T_g$, and follows the Vogel-Fulcher-Tammann relation.

It is evident from (\ref{7}) that the linear susceptibility, $\chi_L =\partial \langle\phi\rangle/\partial h\sim \mu^{-2}=g^2/4I_0v$ ($h$ is an external source of the field $\phi $), is finite in $T_g$. The correlation length, $r_{cor}\sim \sqrt{g^2/4I_0v}$, is finite too. However, nonlinear susceptibility, $\chi_N=\partial^3\langle \phi\rangle/\partial h^3$~\cite{Spin glasses} , diverges near $T_g$:
it is not difficult to check that the nonlinear susceptibility (see Fig.\,4) encloses infinitely increasing contributions. The simplest one of them corresponds to the $h$ diagram in Fig.\,3, and gives the divergent contribution proportional to $(e^{\xi})^{-3!/(2+3\varepsilon/2)}$. Hence in one loop approximation it is possible to estimate $\chi_N\propto (T-T_g)^{-\gamma}$ for $T\to T^{+}_g$, where $12/7< \gamma < 3$. The lower limit is determined by the critical dimension, $d=4$, the upper limit by the real dimension, $d=3$. This is in good agreement with the experimental observations \cite{Spin glasses}.
Thus, one can assert that the system freezes in the state with a disordered structure of the local magnetization field, and $T_g$ is the {\it glass transition temperature}.
\begin{figure}[h]
\label{fig4}
   \centering
   \includegraphics[scale=0.45]{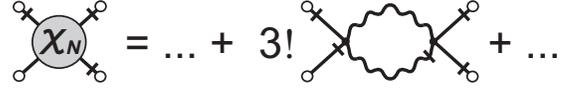}
   \caption{Graphic representation of the nonlinear susceptibility, $\chi_{N}=\langle\phi\bar{\phi}\bar{\phi}\bar{\phi}\rangle_{k=0}$ (the external momenta are $k=0$ in the momentum representation)}
\end{figure}

In conclusion it is necessary to note once again, that the gauge symmetry breakdown is not spontaneous in the considered model. It is induced by the random sources of the gauge field, which is connected with the topological frustrations. The symmetry breakdown leads to the appearance of the gauge field mass. However, the gauge field mass becomes small when the temperature reaches $T_g$ ($T_g>T_c$). Then the scale of the gauge field fluctuations becomes commensurable with the scale of the inhomogeneities induced by the frustrations. Close to $T_g$ the local magnetization fluctuations are small, but the gauge field fluctuations dramatically increase, which leads to the critical slowing-down of the gauge field dynamics, as well as the dynamics of the local magnetization field. This slowing-down freezes the disordered structure of the local magnetization field with the finite correlation length and susceptibility. If frustrations are absent, $I_0\to 0$, the freezing temperature coincides with the phase transition temperature, and the scales of the ordered domains become macroscopic. Then the diagrams with the loops of $\phi$ and $\bar\phi$ fields become divergent, and the system experiences the paramagnetic--ferromagnetic phase transition, which is described within the standard critical dynamics~\cite{CritDyn}.

It should also be noted, that it is not only quenched randomness that leads to the appearance of the gauge field mass.
The disorder in the system can also be induced by the topological frustrations. The geometrical frustrations connected with the topological features of the condensed matter structure can produce a massive gap of $A^a_{\mu }$. For example, in the model of the defected states of the orientation order~\cite{Nelson} the mass of the gauge field is determined by the summed topological charge, $Q$, which is connected with the curvature of the hypersphere, corresponding to the locally preferable structure.

I thank V. N. Ryzhov, V. G. Lebedev and N. M. Chtchelkatchev for stimulating discussions. This work was supported partly by the RFBR grants No. 10-02-00882-a and No. 10-02-00700-a.


\begin{thebibliography}{99}

\bibitem{N1} S.A. Kivelson and G. Tarjus,  Nature Materials, v.7, 831--833(2008);
\bibitem{N2} G.B. Mc Kenna,  Nature Physics, v.4,  673--674, (2008);
\bibitem{Tanaka} H. Tanaka, T. Kawasaki, H. Shintani and K. Watanabe,  Nature Materials, Advance Online Publ., pp.1--8 (2010);
\bibitem{Toul} G. Toulouse, Commun. Phys. 2, 115 (1977);
\bibitem{Hertz} J.A. Hertz,  Phys.Rev.B, {\bf 18,} 4875--4885 (1978);
\bibitem{Volovik} G.E. Volovik, I.E. Dzyaloshinskii,  Sov. Phys. JETP 48(3), 555-559 (1978);
\bibitem{Kanazawa} I. Kanazawa, Journal of Non-Crystalline Solids 293-295, 615--619 (2001);
\bibitem{Riv} N. Rivier, D.M. Duffy,  J.Physique, {\bf 43,} 295--306 (1982);
\bibitem{Nelson} D.R. Nelson,  Phys. Rev. B {\bf 28}, 5515--5535 (1983);
\bibitem{Kivelson} D. Kivelson, G. Tarjus,  Phyl.Mag.B, {\bf 77}, 245--256 (1998);
\bibitem{Nusinov} Zohar Nusinov, Phys. Rev. B {\bf 69}, 014208-1--25 (2004);
\bibitem{Kadich} A. Kadic, D.G.B. Edelen, A gauge theory of dislocations and disclinations, Springer-Verlag, Berlin--Heidelberg--New York, 1983;
\bibitem{Rizhev} N.M. Chtchelkatchev, V.N. Ryzhov, T.I. Schelkacheva, and
E. E. Tareyeva,  Phys. Lett. A {\bf 329}, 244-249 (2004);
\bibitem{Kamenev} A. Kamenev, in Nanophysics: Coherence and Transport, edited by H. Bouchiat, \textit{et al}, [Elsevier, Amsterdam, 2005].
\bibitem{Collins} John C. Collins Renormalization. An introduction to renormalization, the renormalization group, and the operator-product expansion, Cambridge university press, 448 pp., ISBN 0-521-31177-2 (1984).
\bibitem{Vasin} M.G. Vasin, N.M. Shchelkachev, and V.M. Vinokur, Theoretical and Mathematical Physics, 163(1): 537--548 (2010);
\bibitem{Spin glasses}  K. Binder, A.P. Young,  Reviews of Modern Physics, Vol.58, No.4, 801--976 (1986).
\bibitem{CritDyn} C. Hohenberg and B.I. Halperin, Rev. Mod. Phys. 49, 435 (1977);
\end{thebibliography}
\end{document}